\documentclass[prl, letterpaper, amsmath, amssymb, preprintnumbers, showpacs, superscriptaddress, twocolumn, floatfix]{revtex4-1}
\usepackage{graphicx}
\usepackage[usenames]{color}
\pdfoutput=1 

\newcommand{\Tr}{\ensuremath{\text{Tr}}}

\renewcommand{\eqref}[1]{(\ref{eq:#1})}

\begin{document}
\title{Supergravity from D0-brane Quantum Mechanics}

\author{E.~Berkowitz}
\affiliation{Nuclear and Chemical Sciences Division, Lawrence Livermore National Laboratory, Livermore, California 94550, USA}
\author{E.~Rinaldi}
\affiliation{Nuclear and Chemical Sciences Division, Lawrence Livermore National Laboratory, Livermore, California 94550, USA}
\author{M.~Hanada}
\affiliation{Stanford Institute for Theoretical Physics, Stanford University, Stanford, CA 94305, USA}
\affiliation{Yukawa Institute for Theoretical Physics, Kyoto University, Kitashirakawa Oiwakecho, Sakyo-ku, Kyoto 606-8502, Japan}
\affiliation{The Hakubi Center for Advanced Research, Kyoto University, Yoshida Ushinomiyacho, Sakyo-ku, Kyoto 606-8501, Japan}
\author{G.~Ishiki}
\affiliation{Center for Integrated Research in Fundamental Science and Engineering (CiRfSE), University of Tsukuba, Tsukuba, Ibaraki 305-8571, Japan}
\affiliation{Graduate School of Pure and Applied Sciences, University of Tsukuba, Tsukuba, Ibaraki 305-8571, Japan}
\author{S.~Shimasaki}
\affiliation{Research and Education Center for Natural Sciences, Keio University, Hiyoshi 4-1-1, Yokohama, Kanagawa 223-8521, Japan}
\affiliation{KEK Theory Center, High Energy Accelerator Research Organization, Tsukuba 305-0801, Japan}
\author{P.~Vranas}
\affiliation{Nuclear and Chemical Sciences Division, Lawrence Livermore National Laboratory, Livermore, California 94550, USA}

\collaboration{Monte Carlo String/M-theory Collaboration (MCSMC)}
\noaffiliation

\date{15 June 2016}

\preprint{LLNL-JRNL-694846, UTHEP-690, YITP-16-74}

\begin{abstract}
The gauge/gravity duality conjecture claims the equivalence between gauge theory and superstring/M-theory.
In particular, the one-dimensional gauge theory of D0-branes and type IIA string theory should agree on properties of hot black holes.
Type IIA superstring theory predicts the leading $N^2$ behavior of the black hole internal energy to be $E/N^2=a_0T^{14/5}+ a_1T^{23/5}+a_2T^{29/5}+\cdots$ with the supergravity prediction $a_0=7.41$ and unknown coefficients $a_1$, $a_2$, $\ldots$ associated with stringy corrections.
In order to test this duality we perform a lattice study of the gauge theory and extract a continuum, large-$N$ value of $a_0=7.4\pm 0.5$---the first direct confirmation of the supergravity prediction at finite temperature---and constrain the stringy corrections ($a_1 = −9.7\pm2.2$ and $a_2=5.6\pm1.8$).
We also study the sub-leading $1/N^2$ corrections to the internal energy. 

\end{abstract}


\maketitle

{\bf \textit{Introduction:}}
The gauge/gravity duality conjecture~\cite{Maldacena:1997re} has played a central role in theoretical high energy physics for almost two decades.
If the duality is correct, then superstring theory is described by manifestly unitary supersymmetric gauge theories, which provide us with an important key to solve the black hole information loss paradox.
Furthermore, the duality can translate hard problems in strongly coupled field theories in the large-$N$ limit to easier, classical gravity problems.
Given such interesting consequences, it is of crucial importance to provide evidence that this duality holds, by explicitly solving the gauge theory in a regime where non-perturbative effects are dominant.
In a dynamical setup, e.g. at finite temperature, this is extremely difficult.
It is well known that Monte Carlo calculations, analogous to the ones of lattice QCD, are the best tool which can accomplish this task, and provide accurate and improvable results.

Historically, it had been widely believed that the Monte Carlo approach does not work for supersymmetric gauge theories.
The situation has changed in the last fifteen years; various supersymmetric theories relevant for the gauge/gravity duality can now be studied~\cite{Hanada:2016jok}.
However, the calculation can be very expensive.
In this paper we concentrate on the gauge theory of D0-brane quantum mechanics~\cite{deWit:1988wri,Witten:1995im,Banks:1996vh,Itzhaki:1998dd}---this is still expensive, but it is possible to take the continuum and large-$N$ limit using state-of-the-art simulation techniques and supercomputers.

D0-brane quantum mechanics is defined on a Euclidean circle with circumference $\beta$.
With antiperiodic boundary conditions for the fermions and periodic boundary conditions for the bosons, $\beta$ is identified with the inverse temperature $1/T$ of the system.
This model consists of nine $N\times N$ bosonic hermitian matrices $X_M$ ($M=1,2,\cdots,9$), sixteen fermionic matrices $\psi_\alpha$ ($\alpha=1,2,\cdots,16$) and the gauge field $A_t$. 
Both $X_M$ and $\psi_\alpha$ are in the adjoint representation of the $U(N)$ gauge group.
The covariant derivative acts as 
$D_t\cdot = \partial_t \cdot +i[A_t,\cdot]$. 
The continuum Euclidean action is given by $S=S_b+S_f$, where the bosonic part $S_b$ and the fermionic part $S_f$ are given by
\begin{align}
  S_b &= \frac{N}{\lambda}\int_0^\beta dt\ \Tr \left\{ \frac{1}{2}(D_t X_M)^2        - \frac{1}{4}[X_M,X_N]^2       \right\},   \\
  S_f &= \frac{N}{\lambda}\int_0^\beta dt\ \Tr \left\{ i\bar{\psi}\gamma^{10}D_t\psi - \bar{\psi}\gamma^M[X_M,\psi] \right\}.   
\end{align}
while $\gamma^M$ ($M=1,\cdots,10$) are $16\times 16$ the left-handed part of the (9+1)-dimensional gamma matrices.

This model is obtained by dimensionally reducing the ten-dimensional ${\cal N}=1$ super Yang--Mills theory to one dimension.
The index $\alpha$ of the fermionic matrices $\psi_\alpha$ corresponds to the spinor index in ten dimension, and $\psi_\alpha$ is Majorana-Weyl in the ten-dimensional sense.
The 't~Hooft coupling $\lambda$ is related to the Yang--Mills coupling by $\lambda = g_{YM}^2N$.
It has the dimension of $({\rm mass})^3$, and sets the typical energy scale of the theory.
All dimensionful quantities are measured in units of $\lambda$---the dimensionless effective temperature and internal energy are $\lambda^{-1/3}T$ and $\lambda^{-1/3}E$, respectively.
The 't~Hooft limit is $N\to\infty$ with $\lambda^{-1/3}T$ fixed, and $\lambda^{-1/3}E$ scales as $N^2$ there.
In the following, we set $\lambda=1$ for simplicity and without loss of generality.

According to the gauge/gravity duality conjecture, the internal energy in D0-brane quantum mechanics should agree with the mass of the black zero-brane in type IIA superstring theory~\cite{Itzhaki:1998dd} 
\begin{widetext}
\begin{equation}\label{eq:sugra-expectation}
    \frac{E}{N^2} 
    = 
    \frac{a_0 T^{14/5} + a_1 T^{23/5} + a_2 T^{29/5} + \cdots}{N^0} 
  + \frac{b_0 T^{2/5} + b_1 T^{11/5}+\cdots}{N^2}
  + \mathcal{O}(\frac{1}{N^4})
    =
    \frac{E_0(T)}{N^0}
  + \frac{E_1(T)}{N^2}
  + \mathcal{O}(\frac{1}{N^4}).
\end{equation}
\end{widetext}
The leading term $a_0 T^{14/5}$, with $a_0=7.41$, is determined by supergravity. 
Other terms are stringy $\alpha'$- and $g_s$-corrections due to finite string length and virtual string loops, respectively, with $\alpha'\sim T^{3/5}$ and $g_s\sim N^{-2}T^{-21/5}$.
The first term in the $O(N^{-2})$ sector is known to be $b_0=-5.77$ based on an analytic study~\cite{Hyakutake:2013vwa}.

D0-brane quantum mechanics has been investigated using Monte Carlo methods starting with Ref.~\cite{Anagnostopoulos:2007fw}. 
Although existing results suggest a consistency with those expected from the supergravity, the simulations used in these previous tests of the duality were not extrapolated to the continuum limit {\it and} the $N \to\infty $ limit---both these limits are of paramount importance to confirm the duality.
In particular, the results were not precise enough to confirm the coefficient $a_0=7.41$ predicted by supergravity (SUGRA).
In order to obtain this precision, the discretization errors and corrections due to finite $N$ need to be correctly estimated. 
This is achieved for the first time in our study.
Moreover, the high accuracy of our large-scale numerical simulations allows us to robustly determine the first $\alpha'$ correction, resolving a slight tension present in previous studies, and to estimate quantum string corrections.

{\bf \textit{Lattice setup:}} 
To compute observables in D0-brane quantum mechanics using the path integral formulation, we discretize the $0+1$-dimensional spacetime on a linear lattice with $L$ sites.
The length of the circle is $\beta=aL$, where $a$ is the lattice spacing. 
For numerical efficiency, we adopt the static diagonal gauge~\cite{Hanada:2007ti}, 
\begin{equation}
A_t=\frac{1}{\beta}\cdot{\rm diag}(\alpha_1,\cdots,\alpha_N),
\qquad
-\pi<\alpha_i\le\pi. 
\end{equation} 
The corresponding Faddeev-Popov term
\begin{equation}
S_{F.P.}
=- \sum_{i<j}2\log\left|\sin\left(\frac{\alpha_i-\alpha_j}{2}\right)\right|
\label{eq:Faddeev-Popov}
\end{equation}
is added to the action to compensate for the gauge-fixing.

Our lattice action is $S_{F.P.}+S_b+S_f$ where
\begin{eqnarray}
S_b
&=&
 \frac{N}{2a}\sum_{t,M}\Tr \left\{\left(D_+X_M(t)\right)^2\right\}
 \nonumber\\
&&
        -
        \frac{Na}{4}\sum_{t,M,N}\Tr \left\{ [X_M(t),X_N(t)]^2 \right\},
\end{eqnarray}
\begin{eqnarray}
S_f
    &=& 
        \sum_{t}\Tr \Bigl\{iN\bar{\psi}(t)
            \left(
            \begin{array}{cc}
            0 & D_+\\
            D_- & 0
            \end{array}
            \right)
            \psi(t)
            \nonumber\\
            &&
        -
        aN\sum_{t,M}\bar{\psi}(t)\gamma^M[X_M(t),\psi(t)] \Bigl\},  
\end{eqnarray}
where the $t$-independent gauge links are $U=\exp(i a A_t)$.
We improve the covariant lattice derivative so that it is related to the derivative in the continuum theory by $D_\pm\psi(t)= aD_t\psi(t) +\mathcal{O}(a^3)$~\cite{MCSMC2016}.

At finite lattice spacing the theory loses most of its symmetries.
In particular, supersymmetry is broken by the finite lattice spacing and by the boundary conditions, but it is recovered in the continuum limit.

We simulate this theory with the RHMC algorithm with MPI parallelization~\cite{simulation_code}. 
We have neglected the complex phase of the Pfaffian, by using the phase-quenched approximation.
For an argument justifying this procedure, see the longer companion paper Ref.~\cite{MCSMC2016}.

\begin{figure*}[htb]
    \centering
        \includegraphics[width=0.9\textwidth]{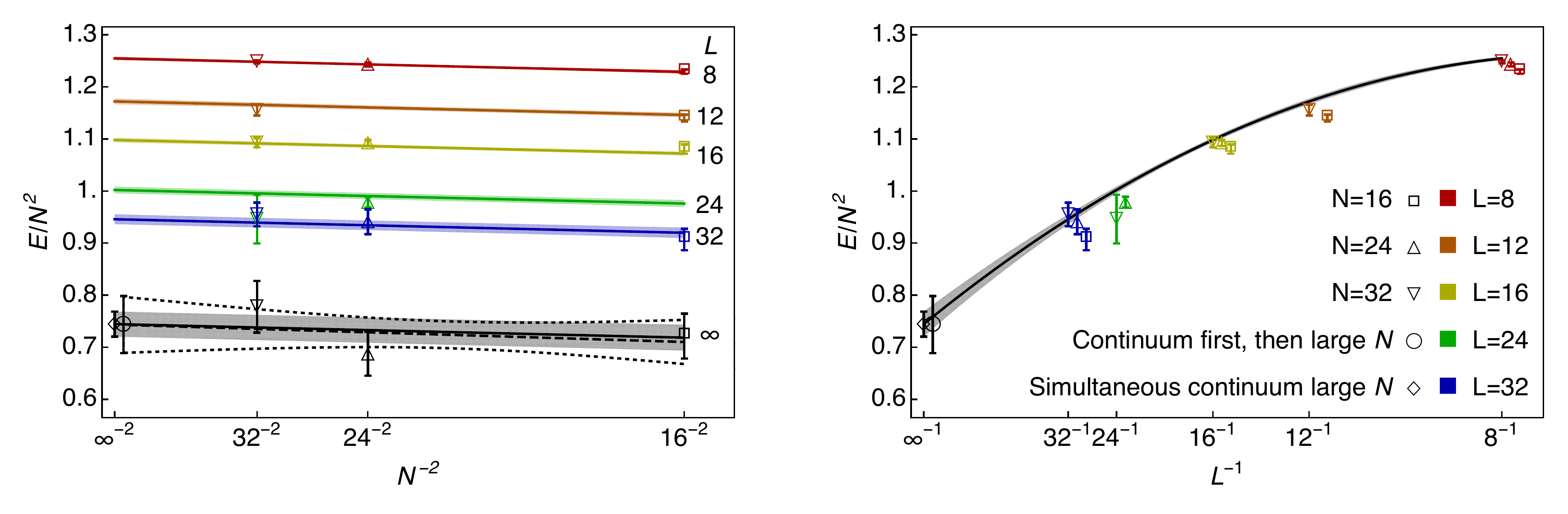}
    \caption{
    A simultaneous continuum- and large-$N$ extrapolation for $T=0.5$ via the surface given by Eq.~\eqref{2D_fit_form}.
    In the right panel, we show all the data points and the $N=\infty$ slice of the best-fit surface.
    The black diamond represents the continuum and large-$N$ corner of the best-fit surface, $e_{00}$.
    The black circle is the result of first performing a continuum extrapolation at each $N$ followed by an extrapolation to large-$N$.
    The continuum extrapolations at each $N$ are shown as black symbols in the left panel.
    We show the large-$N$ extrapolation of those values as a dashed line with dotted bands.
    We also show fixed-$L$ slices of the best-fit surface as solid lines with error bands.
    }
    \label{fig:extrap}
\end{figure*}

We have studied $N=16,24$ and $32$, $T=0.4$ to $1.0$ in steps of $0.1$, with lattice size $L=8, 12, 16, 24$ and $32$.
At each $T$, we performed two kinds of extrapolations, (i) $L\to\infty$ first via a quadratic extrapolation in $L^{-1}$ and a subsequent $N\to\infty$ via a linear extrapolation in $N^{-2}$, and (ii) $L\to\infty$ and $N\to\infty$ together, via
\begin{equation}\label{eq:2D_fit_form}
    \frac{E}{N^2}= e_{00} + \frac{e_{01}}{L} + \frac{e_{02}}{L^2} + \frac{e_{10}}{N^2}.
\end{equation}
The central values of the continuum large-$N$ energy $e_{00}$ are consistent across the two procedures.
An example extrapolation is shown in Fig.~\ref{fig:extrap}.
Henceforth we discuss the results of (ii), which has systematically smaller uncertainties.
For more details about the lattice setup and the continuum large-$N$ extrapolations see Ref.~\cite{MCSMC2016}.

\begin{figure}[htb]
    \centering
        \includegraphics[width=0.45\textwidth]{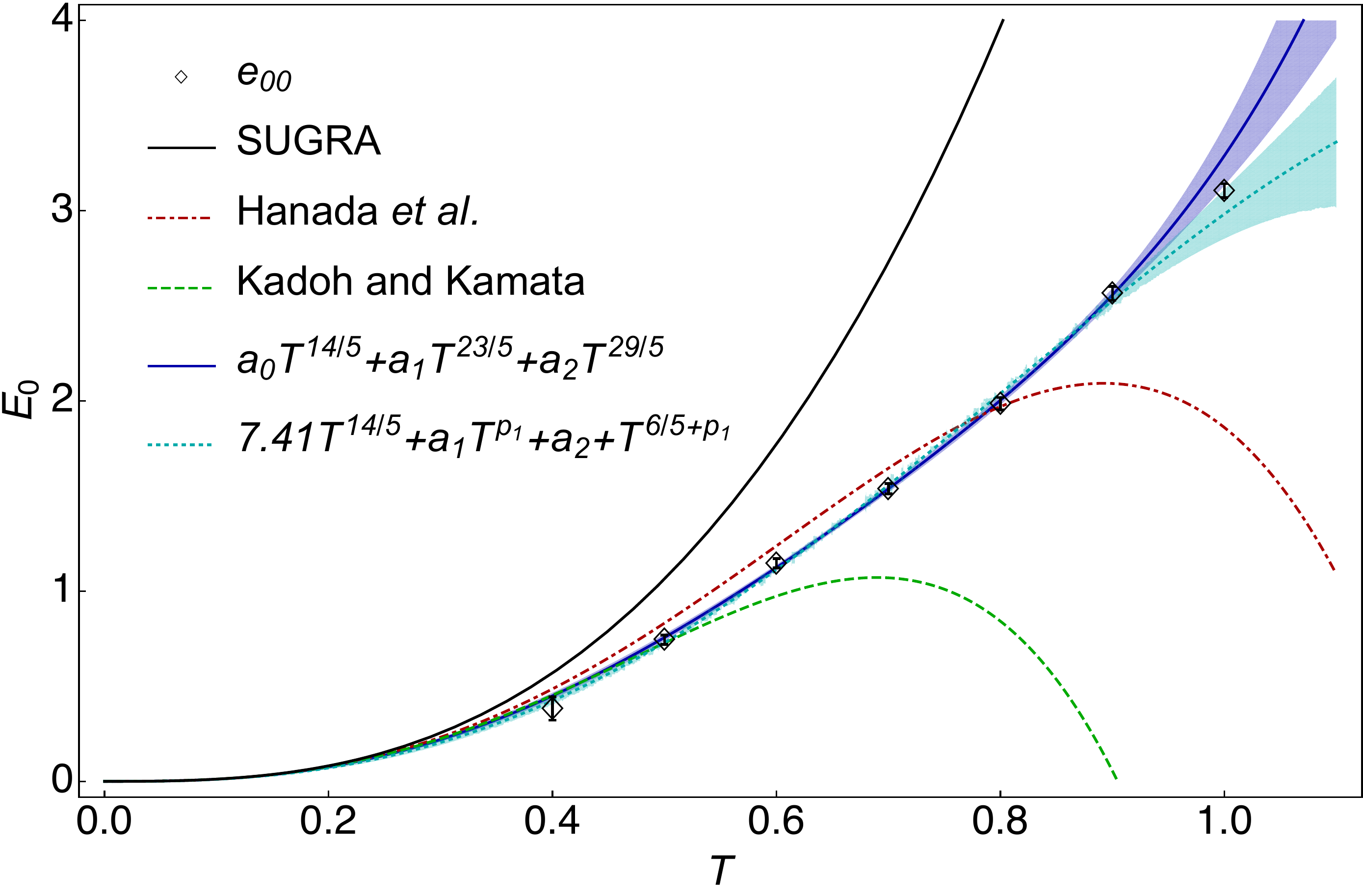}
    \caption{
    Our large-$N$ continuum data $e_{00}$ are shown as black diamonds.
    The solid blue/dotted cyan lines are different fit forms for $E_0(T)$ described in the text.
    We also show the results from Ref.~\cite{Hanada:2008ez}/Ref.~\cite{Kadoh:2015mka} as red dot-dashed/green dashed line.
    The SUGRA result is in black.
    }
    \label{fig:sugra}
\end{figure}

{\bf \textit{Large-$N$ results:}} 
The $N=\infty, L=\infty$ extrapolated values of the energy $e_{00}$ coming from the fit of the measurements of $E/N^2$ at fixed temperature to Eq.~\eqref{2D_fit_form} are shown in Fig.~\ref{fig:sugra}.
The $1/N^2$ corrections in the continuum limit $e_{10}$ are also obtained from the fit.
Our results are the first of this kind: no other numerical study of D0-brane quantum mechanics has ever computed both the continuum limit and the large-$N$ limit.

To test the gauge/gravity correspondence of D0-brane quantum mechanics and supergravity, we want to be able to reproduce the analytical expectation for the leading order term $E_0(T)$ in Eq.~\eqref{sugra-expectation}.
The high accuracy of our extrapolated results $e_{00}$ at several temperatures allows us to perform this test with great precision.
We do a three-parameter fit to $e_{00}$ using $E_0(T)=a_0T^{14/5}+a_1T^{23/5}+a_2T^{29/5}$, scanning different ranges of temperature $0.4\le T\le 0.9$.
Our best fit includes all the data points.
We obtain $a_0=7.4\pm 0.5$, $a_1=-9.7\pm 2.2$ and $a_2=5.6\pm 1.8$ with $\chi^2/{\rm DOF}=2.6/3$.
This result very nicely matches the dual gravity theory expectation $a_0=7.41$ and has a very small uncertainty of about $7\%$.

To test the stability of our fit procedure, we set $a_0$ to 7.41 (its known value) and perform a two-parameter fit to $a_1$ and $a_2$.
We obtain $a_1=-10.0\pm 0.4$ and $a_2=5.8\pm 0.5$, in perfect agreement with the previous fit, increasing our confidence in those results.
These values are also consistent with results of a similar fit at finite-$N$~\cite{Hanada:2016zxj}.

In order to compare with existing results for $a_1$, we perform a different fit based on the function $E_0(T)=7.41T^{14/5}+a_1T^{p_1}$.
This would allow us to also predict the next-to-leading temperature behavior $p_1$, which is expected to be $p_1=23/5=4.6$.
Previously, two results at finite $N$ and without a continuum limit provided slightly different values by fitting to this form:
$a_1=-5.55(7)$, $p_1=4.58(3)$~\cite{Hanada:2008ez} and $a_1=-9(2)$, $p_1=4.7(3)$~\cite{Kadoh:2015mka}.
With our data, we cannot fit that form successfully.
However, a fit to $E_0(T)=7.41T^{14/5}+a_1T^{p_1}+a_2T^{p_1+6/5}$, as motivated from string theory, produces $a_1=-10.2\pm2.4$, $a_2=6.2\pm2.6$ and $p_1=4.6\pm0.3$.
Our results indicate that the previous 2$\sigma$ tension on $a_1$ arose because the next $\alpha'$ correction was not taken into account at temperatures where it is important.

{\bf \textit{$1/N^2$ correction:}} 
We also consider the corrections of order $1/N^2$ to the internal energy, which correspond to the quantum effects arising from virtual loops of strings.
The dual gravity calculation predicts the $1/N^2$ functional form to be $E_1(T)=b_0T^{2/5}+b_1T^{11/5}+\cdots$~\cite{Hyakutake:2013vwa}, with $b_0=-5.77$. 
The first term should become dominant at very low temperature. 
This regime $T<0.1$ has been studied with small $N=3,4,5$ finding good agreement with the gravity prediction~\cite{Hanada:2013rga}.

\begin{figure}[hbt]
    \centering
        \includegraphics[width=0.45\textwidth]{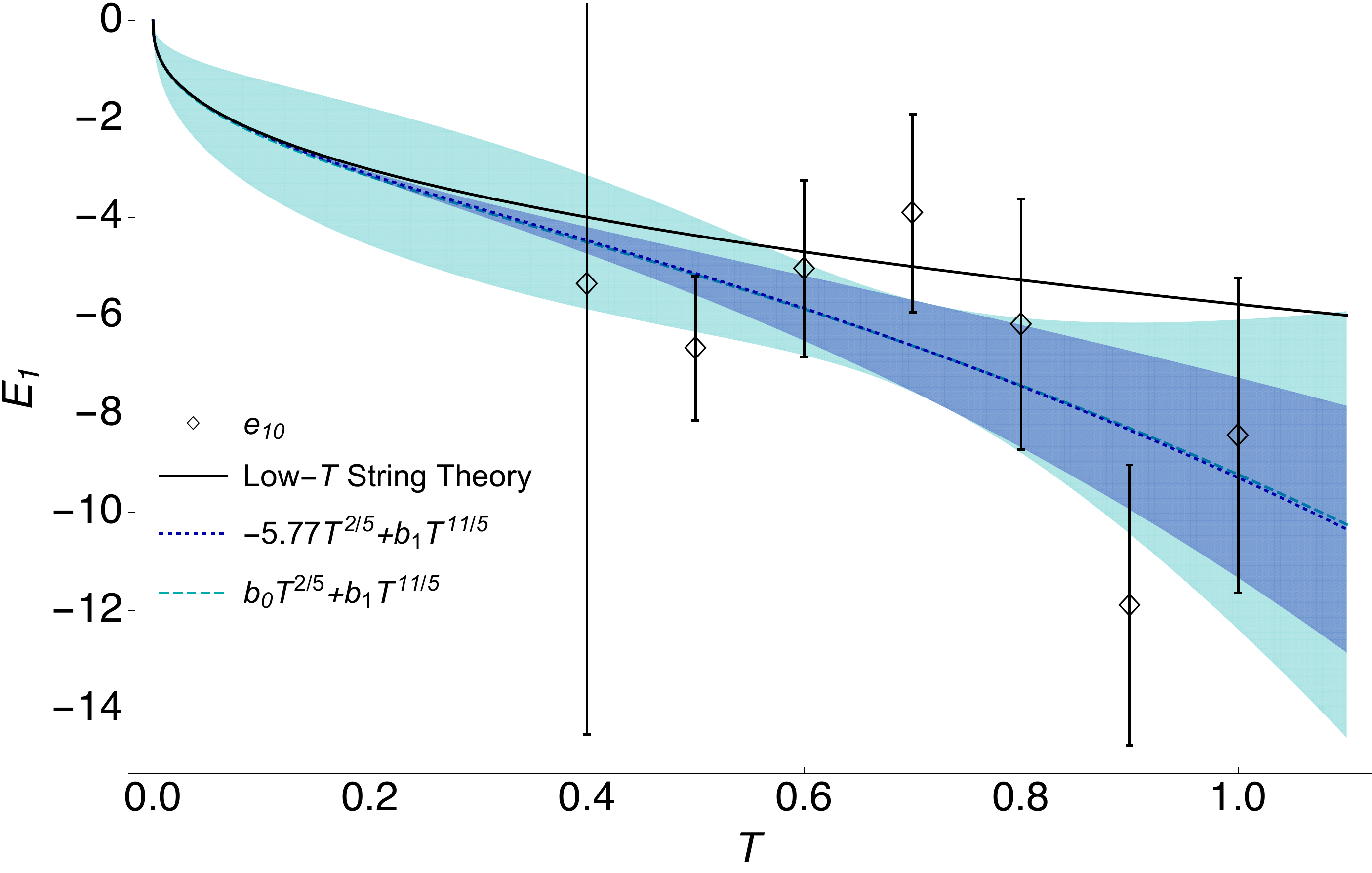}
    \caption{
    Two fits of $E_1(T)$ to our values for $e_{10}$.
    We show our measurements as black diamonds with 1$\sigma$ error bars.
    A fit with fixed/free $b_0$ is shown as a dotted blue/dashed cyan line with a 1$\sigma$ error band.
    The two curves lie on top of one another.
    The known low-temperature behavior $b_0=-5.77$ is shown as a black solid line.
    The one-parameter fit with $b_0$ fixed to be $-5.77$ gives $b_1=-3.5 \pm 2.0$, while a two-parameter fit gives $b_0=-5.8 \pm 3.0$ and $b_1=-3.4 \pm 5.7$.}
    \label{fig:N2}
\end{figure}

Our results at $N=\infty$, where the continuum $1/N^2$ correction ($e_{10}$ in Eq.~\eqref{2D_fit_form}) is extracted directly from the lattice data, are shown as black diamonds in Fig.~\ref{fig:N2}.
We also show the result of a two-parameter fit $b_0T^{2/5}+b_1T^{11/5}$ and a fit of $b_1$ with $b_0$ fixed to its known value.
Although the data is not good enough to extract precision values, a general consistency can be observed.

{\bf \textit{Discussion:}}
The current data suggests that the string $\alpha'$ corrections become negligible at $T\lesssim 0.3$, so that the leading supergravity part, including the power $14/5$, can be determined there.
In this parameter region, the problem of the flat direction becomes more severe~\cite{Anagnostopoulos:2007fw,Hanada:2013rga}, but it should be possible to overcome this difficulty by going to very large $N$ with large-scale parallel simulations. 
Note also that by further improving the precision at $T\gtrsim 0.5$, an accurate test of the quantum ($1/N^2$) string correction is possible, though the low-temperature region studied in Ref.~\cite{Hanada:2013rga} may be more cost-effective.

An even more interesting direction is the study of M-theory.
The black zero-brane is expected to turn to the Schwarzschild black hole in M-theory at very low temperatures, where the temperature scales as a negative power of $N$
\cite{Banks:1996vh,Itzhaki:1998dd}. Also, the plane-wave matrix model \cite{Berenstein:2002jq}, which is a supersymmetric deformation of the D0-brane quantum mechanics, 
is conjectured to describe $M2$- and $M5$-branes \cite{Maldacena:2002rb}.
Large-scale lattice simulation is the only practical tool to verify these conjectures and reveal the dynamical properties of M-theory. 
We believe that strengthening the connection between string theory and lattice gauge theory is an important key for furthering the study of superstring/M-theory and quantum gravity.

\vspace{12 pt}
\noindent {\sc Acknowledgments:}~Numerical calculations were performed on the Vulcan BlueGene/Q at LLNL, supported by the LLNL Multiprogrammatic and Institutional Computing program through a Tier 1 Grand Challenge award, and on the RIKEN K supercomputer.
E.B., E.R., and P.V. acknowledge the support of the DOE under contract~{DE-AC52-07NA27344} (LLNL).
The work of M.H. is supported in part by the Grant-in-Aid of the Japanese Ministry of Education, Sciences and Technology, Sports and Culture (MEXT) for Scientific Research (No. 25287046). 
The work of G.I. was supported, in part, by Program to Disseminate Tenure Tracking System, MEXT, Japan and by KAKENHI (16K17679).
S.S. was supported by the MEXT-Supported Program for the Strategic Research Foundation at Private Universities ``Topological Science'' (Grant No. S1511006).

\raggedright

\end{document}